\documentclass[12pt,preprint]{aastex}

\begin{document}

\shorttitle{Spectroscopic Confirmation of the Dragonfish Association}
\shortauthors{Rahman, Moon \& Matzner}

\title{Spectroscopic Confirmation of the Dragonfish Association:
  \\ The Galaxy's Most Luminous OB Association}

 \author{Mubdi Rahman, Dae-Sik Moon, and Christopher D. Matzner}
 \affil{Department of Astronomy \& Astrophysics, University of
   Toronto, 50 St. George Street, Toronto, ON, M5S 3H4}
 \email{rahman@astro.utoronto.ca}

\begin{abstract}

  Young OB associations with masses greater than 10$^4$ M$_{\sun}$
  have been inferred to exist in the Galaxy but have largely evaded
  detection. Recently, a candidate OB association has been identified
  within the most luminous star forming complex in the Galaxy, the
  Dragonfish Nebula. We identify 18 young, massive stars with
  near-infrared spectroscopy from a sample of 50 members within the
  candidate OB association, including 15 O-type, and three Luminous
  Blue Variables or Wolf-Rayet stars. This number matches the expected
  yield of massive stars from the candidate association, confirming
  its existence and ability to power the parent star forming
  complex. These results demonstrate the existence of a 10$^5$
  M$_{\sun}$ OB association, more powerful than any previously known
  in the Galaxy, comparable in mass only to Westerlund 1. Further, the
  results also validate the color selection method used to identify
  the association, adding credence to others discovered in the same
  way.

\end{abstract}


\section{Introduction}

The most luminous and powerful of a galaxy's young stellar
associations contain a significant fraction of its massive stars, and
dominate all collective forms of energetic feedback into the galaxy's
large-scale wind, gaseous disk, and halo \citep{mckee97, dove00}.
While the massive cluster populations of some other galaxies, such as
M83 are well characterized \citep{chandar10}, only the closest couple
kiloparsecs of the Milky Way have been probed in detail
\citep{lada03}.  Across the Galactic plane, severe obscuration by dust
and the high density of field stars have confounded searches for even
the most luminous associations, leaving our knowledge of them
incomplete. About 50 young OB associations with masses greater than
10$^4$ M$_{\sun}$ have been inferred to exist in the Galaxy
\citep{mckee97, larsen09}; however, they have largely evaded
detection.
 
Our view of the Galactic population of \ion{H}{2} regions is
substantially more complete, thanks to radio surveys \citep{caswell87}
and subsequent follow-up studies.  Of the identifiably coherent
structures which emerge from such work, the most luminous is the
so-called Dragonfish Nebula at ($l$,$b$)=($298$,$-0.4$)
\citep{murray10,rahman10}. Located on the far side of the Galaxy, 9.7
kpc through the Galactic disk, its morphology and radio velocities are
indicative of an expanding bubble. Its H-ionizing luminosity of
10$^{51.8}$ photons per second, if it originates from a coherent group
of stars, requires the presence of the most luminous young stellar
cluster or OB association in the Galaxy. The region's distance and
location in the Galactic plane imply a high degree of extinction,
roughly A$_{K} \sim 1$ or A$_{V} \ga 10$.

Using a near-infrared (NIR) color selection method, a candidate for
the central powering source of the Nebula was identified: the
Dragonfish Association \citep{rahman11}. The association is composed
of 406 $\pm$ 102 candidate O stars, consistent with the measured
ionizing luminosity, making it one of the largest single OB
associations or young clusters in the Galaxy (M $\sim 10^5$
M$_{\odot}$). Notably, the candidate association is not projected
towards one of the bright regions of \ion{H}{2} and 8\,$\mu$m emission
in its vicinity, but within an evacuated shell surrounded by such
emission. This suggests that the ionizing source has inflated a bubble
of radius 69 pc within the Galactic disk. The overdensity of sources
being indirect evidence, we seek direct spectroscopic confirmation
that the overdensity consists of O stars, and that the field objects
are K giants.

In this Letter, we confirm the existence of the Dragonfish Association
through NIR spectroscopy of massive stars. In addition, we identify
two candidate Luminous Blue Variables and a Wolf-Rayet within the
association. This verifies the presence of a OB association which, on
the basis of source counts, is sufficiently luminous to power its host
star forming complex, giving it a total mass of 10$^5$ M$_{\sun}$ and
making it the most luminous OB association in the Galaxy. The
confirmation of the Dragonfish association validates the NIR color
selection method as a way of identifying candidate clusters and
associations in distant star forming complexes.

\section{Observations \& Data Reduction}

We conducted H- and K-band spectroscopy (1.52 - 2.4 $\mu$m) of 50
“cluster” sources projected towards the candidate association, and a
control sample of four non-cluster sources, between 2011 March 16 to
20. We use the SOFI spectrograph at the 3.6m ESO New Technology
Telescope with the red grism and 0\farcs6 slit providing a spectral
resolving power of R $\sim$ 1000. Both samples are limited to NIR
colors consistent with extinguished O stars (or less extinguished K
giants), and to K magnitudes brighter than 12. The control sample is
offset from the candidate association by 0\fdg5 and should represent
the interloper sources within the cluster sample. The total
integration time is 60 min for each cluster source and 2 min for each
non-cluster source, using the nod-and-shuffle technique. This provides
minimum signal-to-noise ratios of $> 250$ and $>50$, respectively. The
data was reduced and analyzed with Python, using the {\it numpy} and
{\it scipy} packages.

\section{Analysis \& Results}
\subsection{Spectral Classification}

From the NIR color identification method, the contaminating non-member
population confused with the candidate association is predominantly K
giants with moderate extinction \citep[A$_K \simeq$
0.3;][]{rahman11}. Due to the high extinction, the intervening
interstellar medium will imprint absorption lines on the spectra as
well, primarily from neutral metals (see \S
\ref{subsect:specfeat}). The NIR spectra of K giants are similarly
riddled with neutral metal lines, causing difficulty in identifying O
stars on this basis, at least at low spectral resolution
\citep{rayner09}.

To discriminate between extinguished O stars and K giants, we
concentrate on two absorption lines which are characteristic of
massive young stars but absent or very weak in K giants. These are the
2.166 $\mu$m Br $\gamma$ line and the 1.700 $\mu$m \ion{He}{1} line,
which is prominent in late O stars. A number of weaker features, such
as the 1.693 $\mu$m \ion{He}{2} line (present in early O stars), 1.681
$\mu$m \ion{H}{1} line, and the 2.188 $\mu$m \ion{He}{2} line, are
also useful for identification.

\begin{figure}
\begin{center}
\includegraphics[scale=.75]{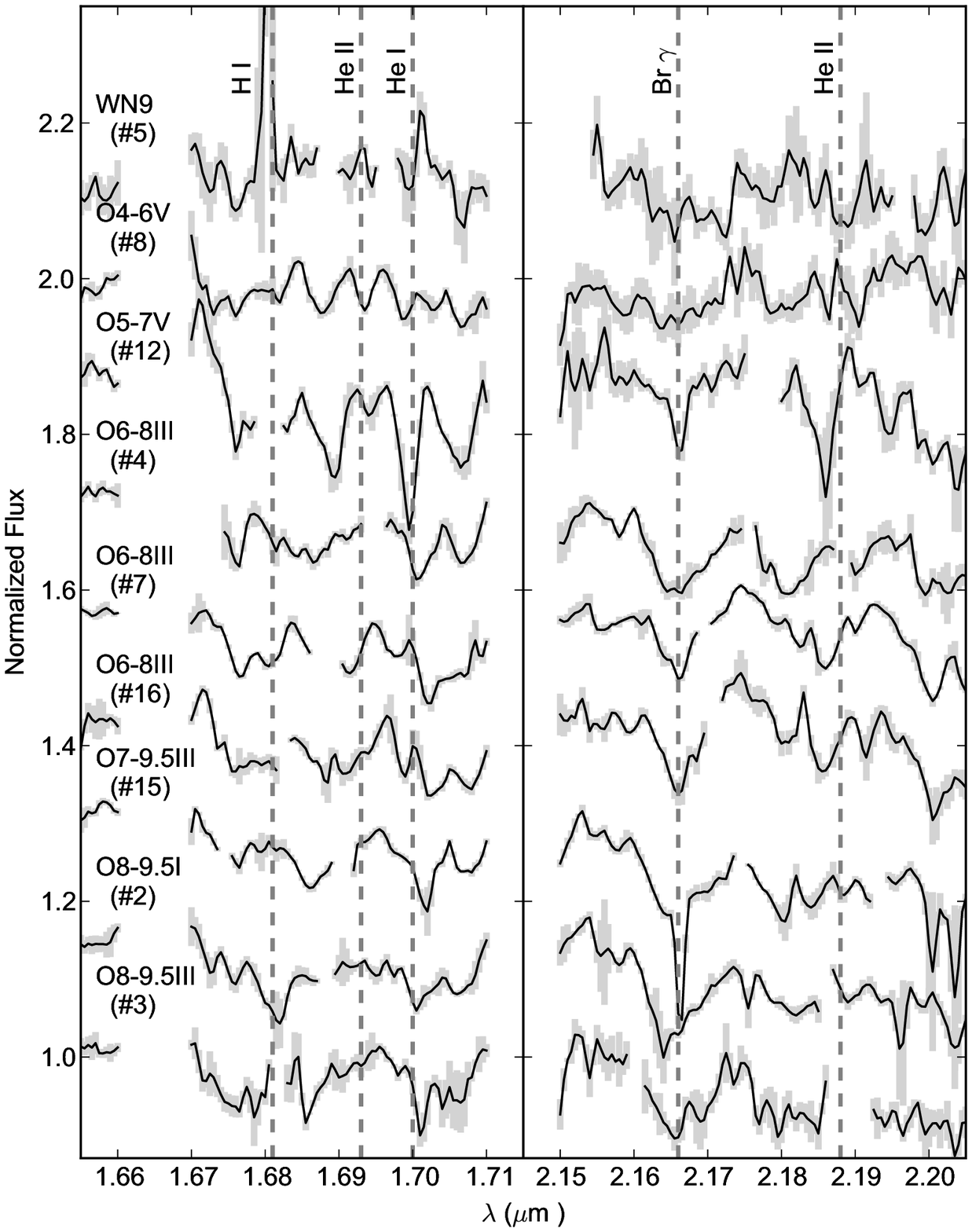}
\caption{The spectra of 9 massive stars within the Dragonfish
  Association at the wavelength ranges of interest for spectral
  classification. The spectral types and identification numbers from
  Table 1 are indicated. The wavelengths of features used to classify
  the stars are indicated with dashed lines. The spectra have not been
  adjusted for Doppler shifts. The error bars are indicated in
  grey. \label{fig:waterfall}}
\end{center}
\end{figure}

In Table 1, we present the identified young massive stars: 15 O stars
and three evolved massive stars. We present the spectra of a selection
of these stars in Fig \ref{fig:waterfall}. We classify the NIR spectra
of O stars with the Brackett series of hydrogen absorption lines and
He I and II absorption lines. These features are weak, necessitating
signal-to-noise ratios above 100 for detection \citep{hanson05}. We
determine a subtype range and luminosity class with the aid of the
weaker absorption lines and NIR photometry using the NIR extinction
law from \citet{nish09} and model photometry from
\citep{martins06}. The magnitude differences between luminosity
classes (from 0.2 to 1.0 mag in the K-band) of the same spectral
subtype ranges make it improbable for small differences in the
extinction determination to change the luminosity class
identification.  We note that massive stars with weaker features, such
as early O stars lacking \ion{He}{1} absorption or with Brackett
$\gamma$ absorption that has been filled in, may evade detection
\citep{hanson05}. All except the most luminous stars (the WN9 and O4-6
stars) show strong Br $\gamma$ absorption. The later stars all show
\ion{He}{1} absorption while the earlier stars show \ion{He}{2}
absorption.

\begin{figure}
\begin{center}
\includegraphics[scale=.7]{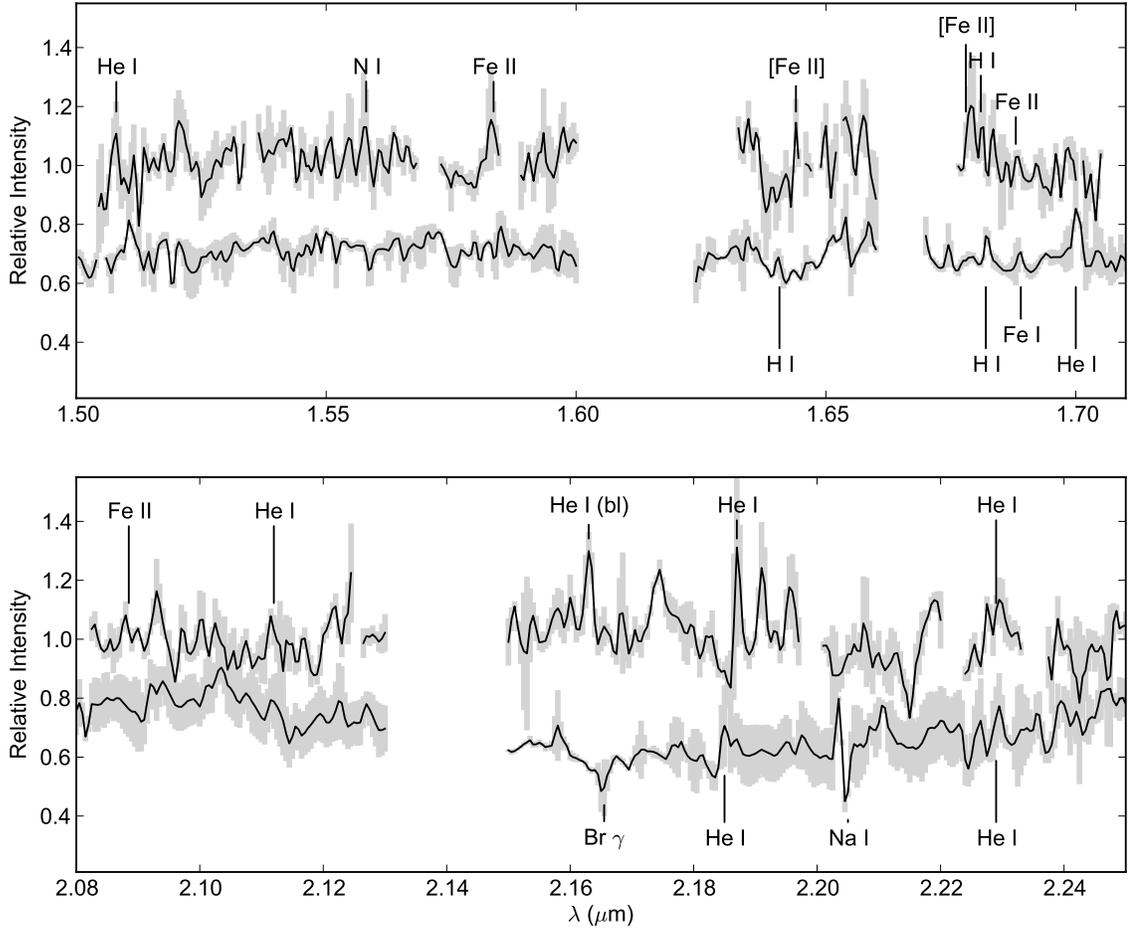}
\caption{The spectra of the two candidate luminous blue variable
  candidates. The upper spectrum is star 9 and the lower is star 10
  from Table 1. The error bars are indicated in grey. Characteristic
  spectral features are indicated. All spectral features are
  significantly stronger than those of the
  O-stars. \label{fig:candlbv} }
\end{center}
\end{figure}

In addition to the 15 O stars, we identify one Wolf-Rayet star based
on \ion{H}{1}, \ion{He}{1} and \ion{He}{2} emission
\citep{morris96}. Further, we find two candidate Luminous Blue
Variables (LBVs), based on numerous He I emission lines, and
[\ion{Fe}{2}] emission in one case
\citep[Fig. \ref{fig:candlbv};][]{morris96}. These rare, exotic
objects, transition stages of the most massive stars, are typically
found in locations of recent star formation. Confirmation of the LBVs
will require studies of their photometric variability.

\subsection{Interstellar Features}
\label{subsect:specfeat}

The high signal-to-noise spectra of the massive stars presented in
figure \ref{fig:waterfall} contain a number of features that are
similar to features in the contaminating K-giant population. Riddled
throughout the spectra are neutral metal lines, primarily from Mg, Al,
Fe, and Si. These elements are common in the interstellar medium and
have been observed in absorption within the local 100 pc
\citep{redfield04}. As such, these lines are expected to be imprinted
on relatively featureless spectra when seen through a large column, as
would be expected towards the association.

In addition to the metal absorption lines, strong CO vibrational
absorption bands are visible in the spectra of all identified massive
stars at 2.29, 2.32 ($\nu$ = 2 $\rightarrow$ 0) and 2.35 ($\nu$ = 3
$\rightarrow$ 1) \micron. These are often detected as photospheric
absorption features of K-giants \citep{rayner09}, however, they may be
also imprinted on nearly featureless spectra of younger, more massive
stars by interstellar absorption. The $\nu$ = 2 $\rightarrow$ 0 bands
have been detected towards NGC 2024 IRS 2 with line widths $\Delta$v =
1.4 km s$^{-1}$ \citep{black84}. The individual lines within each band
are found to be saturated with total band equivalent widths of
$\sim$0.5 \AA{} \citep{lacy94}. Towards the Dragonfish region, there
is CO emission detected between 10 km s$^{-1}$ $<$ v$_{\rm lsr}$ $<$
30 km s$^{-1}$ coincident with the Dragonfish association and its
surrounding photon-dominated region \citep[PDR;][]{grabel88,
  murray11}. In the stacked cluster O-star spectra (see \S
\ref{subsect:stacked}), the equivalent width of each of the CO bands
is $\sim$5 \AA, consistent with what is seen towards NGC 2024 when
saturated over a larger range of velocities.

For the $\nu$ = 3 $\rightarrow$ 1 band to arise in the ISM, sufficient
CO with excitation temperatures exceeding 1000 K must exist to
populate the $\nu$ = 1 vibrational state.  The PDR traced by
polycyclic aromatic hydrocarbon emission in the region
\citep{rahman10}, is a likely site of such warm CO
\citep{hollenbach97}. We note that CO $\nu$ = 2 $\rightarrow$ 0 and
$\nu$ = 3 $\rightarrow$ 1 bands have been observed in absorption
towards massive stars in Cygnus OB2 classified by NIR and optical
spectroscopy, at significantly lower dust extinctions than the
Dragonfish's \citep{comeron08}. Therefore, we conclude that CO
absorption bands in the NIR spectra of the identified O stars are
plausibly interstellar in origin, at least when the O stars are
surrounded by an intensely illuminated PDR. However, this point
deserves additional study.

\subsection{Stacked Spectra Comparison}
\label{subsect:stacked}
\begin{figure}
\begin{center}
\includegraphics[scale=.75]{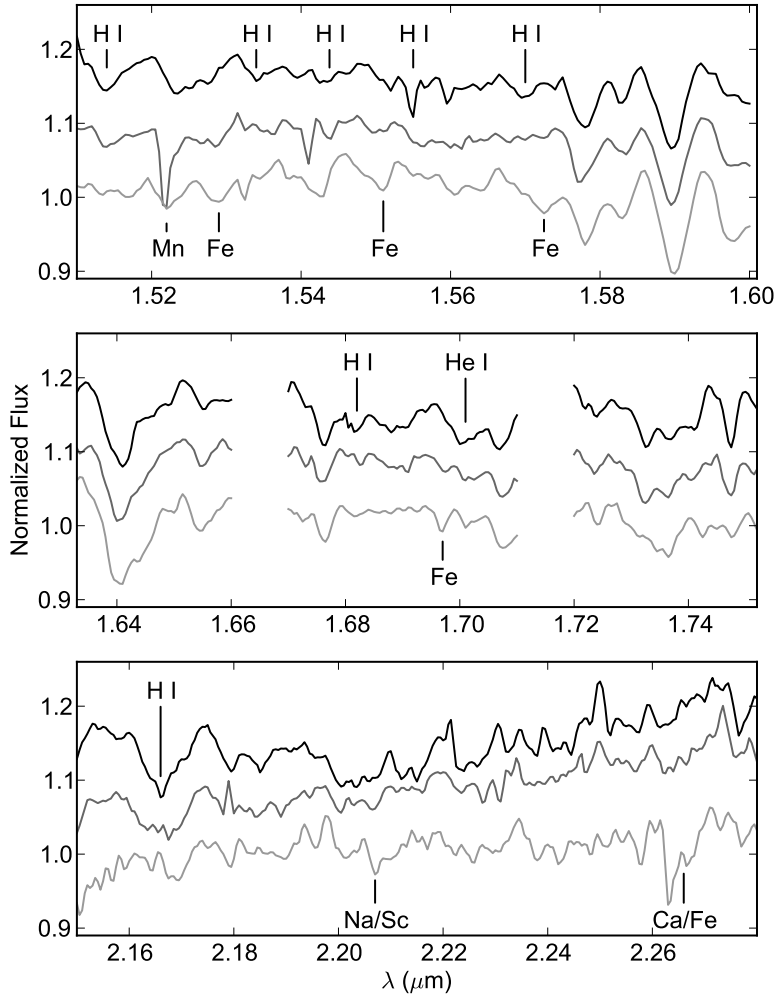}
\caption{The stacked cluster O star (top, black line), cluster
  contaminant (middle, lighter, line) and non-cluster (bottom,
  lightest line) spectra. The error of the stacked spectra is at the
  0.01\% level, and so is not indicated. Characteristic distinguishing
  spectral features are indicated for the O stars and K giants. The
  cluster contaminants likely include massive stars with features too
  weak to be identified on individual spectra. \label{fig:stacked} }
\end{center}
\end{figure}

To investigate the weak NIR features in a more robust manner, we
construct three groups of sources and examine the stacked spectra of
each: the 15 cluster O stars, the 32 cluster sources not immediately
identified as massive stars, and the four non-cluster field sources
from our control sample. We present the stacked spectra in figure
\ref{fig:stacked}. The primary spectral discriminators, Br $\gamma$
and 1.700 $\mu$m \ion{He}{1}, are strong in the cluster O stars.  Br
$\gamma$ is weaker but visible in the cluster contaminant population,
and both lines are absent in the control non-cluster
population. Further, weaker \ion{H}{1} absorption features at 1.514,
1.534, 1.544, 1.556, 1.570 and 1.681 $\mu$m, characteristic of massive
stars rather than cool giants, are detected in the O star population
and absent in the others. Conversely the Ca and Fe feature at 2.26
$\mu$m, a characteristic feature in the spectra of cool stars, is
strong in the non-cluster population, weak in the cluster
contaminants, and absent in the O stars. In addition, differences
between the three populations are visible with lines at 1.522 $\mu$m
Mn, 1.529, 1.551, 1.573, and 1.697 $\mu$m Fe, and 2.207 $\mu$m Na/Sc,
all of which are present in the non-cluster and contaminant
populations, and either weak or absent in the O star population.
These spectral differences confirm that the identification of the
association O stars is robust.

\subsection{Possible Runaway?}

One of the identified O stars, Source 14 in Table 1, appears to be
highly blueshifted in the 1.693 $\mu$m \ion{He}{2}, 1.700 $\mu$m
\ion{He}{1}, 1.736 $\mu$m \ion{H}{1} and Br $\gamma$ absorption
lines. The blueshift of the lines corresponds to a radial velocity of
300 $\pm$ 50 km s$^{-1}$, as compared to the other O stars that show
no signficant Doppler shift. The velocity is much larger than can be
explained by Galactic rotation. This indicates that the star is in a
close binary system, or is a runaway O star either kicked out of the
association or originating elsewhere. In our single-epoch spectrum, we
find no indications of a companion star in the spectrum, such as
splitting of spectral lines. If it is a runaway, it could have been
produced by a binary-binary encounter or by the result of the
supernova explosion of the primary of a massive binary
\citep{gvar08}. Similar massive runaways have been identified
originating from Cygnus OB2 \citep{comeron07}.

\section{Discussion \& Summary}

The spectroscopic identification of massive, young stars confirms the
existence of a young OB association within the Dragonfish Nebula, but
its share of the nebula's ionization budget must be addressed
statistically. Using the extracted candidate cluster properties from
\citet{rahman11}, we infer a membership probability of 35.5 \%, for an
expected yield of 17.8 $\pm$ 3.4 members in the cluster
sample. Indeed, 18 of our cluster sources are young, massive
stars. This confirms that the Dragonfish Association is real, and not
a chance sky overdensity of other stellar types (e.g. K-giants)
coincident with the nebula or a spurious effect of intervening
extinction. Further, this indicates that the NIR color selection
method produces association membership probabilities consistent with
the actual yield within the association.

From the point sources identified using the color selection method,
assuming a minimum candidate spectral type of O9.5V \citep{martins05}
and using a standard stellar initial mass function \citep{kroupa01},
the inferred total stellar mass of the association is $\sim 10^5$
M$_{\sun}$ \citep{rahman11}. This is remarkably consistent with the
estimated mass of the minimum stellar population required to power its
central star forming complex \citep{rahman10}. This indicates that the
Dragonfish Association dominates the ionization of its surrounding
nebula. Further, the mass of this association is greater than other
luminous associations in the Galaxy such as NGC 3603 ($1.3 \times
10^4$ M$_{\odot}$), Trumpler 16 in the Carina Nebula ($1.8 \times
10^4$ M$_{\odot}$), Cygnus OB2 ($7.6 \times 10^4$ M$_{\odot}$), and
the Arches \citep[$7.7 \times 10^4$ M$_{\odot}$;][]{weidner10}. It is
similar to that of Westerlund 1 \citep[10$^{5}$ M$_{\sun}$;
][]{clark05}, however the Dragonfish is significantly more luminous
compared to Westerlund 1 based on its free-free luminosity
\citep[Westerlund 1 is not visible in the WMAP free-free maps;
][]{murray10}. This is consistent with the more evolved state of the
massive stars in the latter. If the Dragonfish association is, in
fact, the sole engine of the free-free emission in the complex, its
output of 10$^{51.8}$ H-ionizing photons per seconds is similar to
R136, the dominant powering source of the most luminous \ion{H}{2}
region in the Local Group, 30 Doradus in the Large Magellanic Cloud
\citep{parker93,evans11}.

The identification of O stars and evolved massive stars confirms the
existence of the Dragonfish Association. The association is not
embedded within a bright \ion{H}{2} region, but rather a void of
continuum and PAH emission surrounded by a shell, indicating that it
is the central illuminating and inflating source of the Galaxy's most
luminous star forming complex. As an extreme example of Milky Way star
formation, the Dragonfish complex merits further study in a number of
ways. Its large population of O stars, an estimated 13 with masses $>
100$ M$_{\sun}$ \citep[using the initial mass function
  from][]{kroupa01}, makes it an ideal laboratory for investigations
of environmental influences at the upper limit of stellar masses. With
high-precision astrometry, its membership can be more firmly
established and its internal kinematics can be probed. X-ray, radio,
and $\gamma$-ray follow-up is warranted to identify colliding wind
binaries and young stellar objects. Searches for IR excess sources
should reveal a population of new protostars triggered by its
influence. Moreover, this influence should be sought in the dynamics
of atomic and molecular gas in its environment.

Additional candidate associations have been identified towards star
forming complexes identified by \citet{rahman10} using this method and
will be characterized in a forthcoming paper. The confirmation of the
Dragonfish Association adds significant credence to the hypothesis
that colour-selected clusters found toward the Galaxy's most luminous
star forming complexes are in fact their driving stellar associations.

\acknowledgments

This work is based on observations collected at the European Southern
Observatory, Chile (ESO 086.C-0502). We thank N. Murray, J. Graham,
P.G. Martin and R. Breton for comments and discussion. D-S.M. and
C.D.M acknowledge support from the Natural Science and Engineering
Research Council of Canada. We acknowledge the staff of the ESO La
Silla Observatory. This paper is dedicated to the memory of Crystal
Marie Brasseur.

{\it Facilities:} \facility{NTT (SOFI)}

\begin{deluxetable}{cccccccc}
\tablewidth{0pt} 

\tablecaption{Massive Stars Identified in the Dragonfish Association}

\tablehead{ \colhead{\#} & \colhead{RA} & \colhead{Dec} & \colhead{l
    (deg)} & \colhead{b (deg)} & \colhead{Spectral Type} &
  \colhead{K$_{S}$ (mag)} & \colhead{A$_{K}$ (mag)} }

\startdata 

1 & 12h11m30.2s & -63\degr15\arcmin35\farcs7 & 298.458 & -0.736 & O6-8I & 10.5 &
1.2 \\

2 & 12h11m32.0s & -63\degr15\arcmin52\farcs7 & 298.462 & -0.740 & O8-9.5I & 10.5 &
1.1 \\

3 & 12h11m41.4s & -63\degr16\arcmin31\farcs4 & 298.481 & -0.748 & O8-9.5III & 12.0
& 1.0 \\ 

4 & 12h11m48.2s & -63\degr14\arcmin33\farcs6 & 298.488 & -0.713 & O6-8III & 11.1 &
1.0 \\ 

5 & 12h11m54.1s & -63\degr17\arcmin04\farcs0 & 298.506 & -0.753 & WN9 & 11.5 & 1.2
\\

6 & 12h11m58.4s & -63\degr09\arcmin42\farcs1 & 298.495 & -0.630 & O8-9.5III &
11.6 & 0.9 \\ 

7 & 12h12m06.5s & -63\degr11\arcmin36\farcs1 & 298.515 & -0.659 & O6-8III & 11.0 &
1.1 \\ 

8 & 12h12m11.1s & -63\degr12\arcmin20\farcs8 & 298.525 & -0.670 & O4-6V & 11.6 &
1.1 \\ 

9 & 12h12m20.3s & -63\degr13\arcmin42\farcs5 & 298.546 & -0.690 & LBV? &
11.9 & 1.1 \\ 

10 & 12h12m21.4s & -63\degr16\arcmin38\farcs4 & 298.555 & -0.738 & LBV? & 9.0 &
1.1 \\ 

11 & 12h12m22.1s & -63\degr13\arcmin40\farcs9 & 298.549 & -0.689 & O9-9.5III &
11.7 & 1.0 \\ 

12 & 12h12m29.5s & -63\degr10\arcmin51\farcs5 & 298.556 & -0.641 & O5-7V & 11.9 &
1.1 \\ 

13 & 12h12m37.2s & -63\degr14\arcmin52\farcs5 & 298.580 & -0.705 & O4-6V & 11.6 &
1.1 \\ 

14 & 12h12m41.3s & -63\degr09\arcmin27\farcs3 & 298.574 & -0.614 & $<$O5V & 11.9 &
1.1 \\ 

15 & 12h12m52.3s & -63\degr17\arcmin03\farcs9 & 298.613 & -0.736 & O7-9.5III &
11.2 & 1.0 \\ 

16 & 12h12m52.6s & -63\degr18\arcmin30\farcs8 & 298.618 & -0.760 & O6-8III & 11.4
& 1.0 \\ 

17 & 12h12m54.4s & -63\degr17\arcmin56\farcs5 & 298.619 & -0.750 & O6-8I & 11.1 &
1.2  \\ 

18 & 12h12m55.3s & -63\degr06\arcmin40\farcs6 & 298.593 & -0.564 & O9-9.5III &
11.3 & 1.0 \\ \enddata

\end{deluxetable}

\end{document}